\documentclass[12pt]{article}
\usepackage{amssymb}
\newtheorem{proposition}{Proposition}[section]  %proposition
\newcommand{\bprop}{\medskip\begin{proposition} ~~\\ \it}
\newcommand{\eprop}{\end{proposition} \hfill $\Box$ }% \bigskip}
\newtheorem{naming}{Definition}[section]   %definition
\newcommand{\bdefi}{\medskip\begin{naming} ~~\\ \it}
\newcommand{\edefi}{\end{naming} \hfill $\Diamond$ }% \bigskip}
\newtheorem{example}{Example}[section]   %example
\def\bexam{\medskip\begin{example} ~~\\ \rm}
\def\eexam{\end{example} \hfill $\triangle$ }% \bigskip}
\newcommand{\sect}[1]{\setcounter{equation}{0}\section{#1}}
\newcommand{\subsect}[1]{\subsection{#1}}

\newcommand{\be}{\begin{equation}}
\newcommand{\ee}{\end{equation}}
\newcommand{\bea}{\begin{eqnarray}}
\newcommand{\eea}{\end{eqnarray}}
\newcommand{\bean}{\begin{eqnarray*}}
\newcommand{\eean}{\end{eqnarray*}}
\newcommand{\nn}{\nonumber}
\newcommand\IC{{\mathbb C}}

\newcommand\IF{{\mathbb F}}
\newcommand\IH{{\mathbb H}}

\newcommand\IM{{\mathbb M}}

\newcommand\IR{{\mathbb R}}

\def\abs#1{{\vert#1\vert}}

\def\inf1{{\cal L}^{(1,\infty)}}

\def\otr{\otimes_{\IR}}

\def\bar#1{\overline{#1}}

\def\bra#1{\left\langle #1\right|}
\def\ket#1{\left| #1\right\rangle}
\def\hs#1#2{\left\langle #1|#2\right\rangle}

\def\ca{{\cal A}}
\def\cb{{\cal B}}

\def\cg{{\cal G}}

\def\cu{{\cal U}}

\def\raw{\rightarrow}

\def\lrw{\leftrightarrow}
\begin{document}
\setcounter{page}{0}
\thispagestyle{empty}
\begin{flushright}
\hfill Trieste-DSM-QM442 \\
\hfill December 1998
\end{flushright}
\vspace{.5cm}
\begin{center}{\Large \bf Deconstructing Monopoles and Instantons}
\end{center} 
\vspace{1cm}
\centerline{\large  Giovanni Landi}
\vspace{5mm}
\begin{center}
Dipartimento di Scienze Matematiche,
Universit\`a di Trieste \\ P.le Europa 1, I-34127, Trieste, Italy 
\\ and INFN, Sezione di Napoli, Napoli, Italy. 
\end{center}
\vspace{2.5cm}
\begin{abstract}
We give a unifying description of the Dirac monopole on the $2$-sphere $S^2$, of a
graded monopole on a $(2,2)$-supersphere $S^{2,2}$ and of the BPST instanton on the
$4$-sphere $S^4$, by constructing a suitable global projector $p$ via equivariant
maps. This projector determines the projective module of finite type of
sections of the corresponding vector bundle. The canonical connection 
$\nabla = p \circ d$ is used to compute the topological charge which is found to
be equal to $-1$ for  the three cases. The transposed projector $q=p^t$ gives the
value $+1$ for the charges; this showing that transposition of projectors, although
an isomorphism in $K$-theory, is not the identity map. We also study the
invariance under the action of suitable Lie groups.
\end{abstract}

\vfill
{\hfill \it This work is dedicated to Matteo}

\newpage
\sect{Preliminaries and Introduction}
It is well known since the early sixties that vector bundles can be though of as
projective modules of finite type (finite projective modules `for short'). The
Serre-Swan's theorem \cite{Sw} states that there is a complete equivalence between
the category of (smooth) vector bundles over a (smooth) compact manifold $M$ and
bundle maps, and the category of finite projective modules over the commutative
algebra $C(M)$ of (smooth) functions over $M$ and module morphisms. The space
$\Gamma(M,E)$ of smooth sections of a vector bundle $E \raw M$ over a compact
manifold $M$ is a finite projective module over the commutative algebra $C(M)$ and
every finite projective $C(M)$-module can be realized as the module of sections of
some vector bundle over $M$. In fact, in \cite{Sw} the correspondence is stated in
the continuous category, meaning for topological manifolds and vector bundles and
for functions and sections which are continuous. However, the equivalence can be
extended to the smooth case \cite{Co1}.
This correspondence was already used in \cite{Ko} to give an algebraic version of
classical geometry, notably of the notions of connection and covariant derivative.
But it has been with the advent of noncommutative geometry \cite{Co} that the
equivalence has received a new emphasis  and has been used, among several other
things, to generalize the concept of vector bundles to noncommutative geometry and
to construct noncommutative gauge and gravity theories. Furthermore, since the
creation of noncommutative geometry, finite projective modules are increasingly
being used  among (mathematical)-physicists. 

In this paper we present a finite-projective-module description of
the basic topologically non trivial gauge configurations, namely monopoles and
instantons. This will be done by constructing a suitable global projector 
$p\in \IM_N(C(M))$, the latter being the algebra of $N\times N$ matrices whose
entries are elements of the algebra $C(M)$ of smooth functions defined over the
base space.  That $p$ is a projector is expressed by the conditions $p^2 = p =
p^\dagger$. The module of sections of the vector bundles on which monopoles or
instantons live is identified with the image of $p$ in the trivial module
$C(M)^N$ (corresponding to the trivial rank $N$-vector bundle over $M$), i.e. as
the right module
$p (C(M))^N$.

Now, not all the projectors that we construct are new.
The Dirac monopole \cite{Di} projector is already present in \cite{Ka} (in fact,
the monopole projector is well know among physicists), while the BPST instanton
\cite{BPST} is present in the ADHM analysis \cite{At}, albeit in a local form.
Our presentation is a global one which does not use any local chart or partition of
unity and it is based on a unifying description in terms of global equivariant
maps. We express the projectors in terms of a more fundamental object, a
vector-valued function of basic equivariant maps. It is this reduction that
has motivated the word `{\it deconstructing}' in our title. 

For the time being, we present only the projectors carrying the lowest values of
the charges, i.e. $\pm 1$. Now, when the sphere $S^2$ is regarded as the complex
projective space $\IC P^1$ or the compactified plane $\IC_\infty = \IC 
\cup \{\infty\}$ the monopole projector translates into the Bott projector (see
for instance \cite{W-O}). Thus, for the sphere $S^4$ and the supersphere
$S^{2,2}$ the projectors we construct could be considered as analogues of the
Bott projector for $S^2$. These three projectors will then give a generator of the
reduced $K$-theory groups \cite{Ka} $\widetilde{K}(S^2)$, $\widetilde{K}(S^{2,2})$
and
$\widetilde{K}(S^4)$ respectively. The construction of global (i.e. without
partition of unity and local charts) projectors for all values of the charges as
well as for projective spaces will be the content of a paper in preparation
\cite{La}. 

We refer to \cite{book} for a friendly approach to modules of several kind
(including finite projective). Throughout the paper we shall avoid writing
explicitly the exterior product symbol for forms.  

\sect{The General Construction}
In this section we shall briefly describe the general scheme that will be
used in the following for the monopoles and the instantons. All
ingredients will be defined explicitly when needed later on. So let
$M$ be the sphere $S^2$, the supersphere $S^{2,2}$, or the sphere $S^4$. The
symbol $G$ will indicated the group $U(1)$, a supergroup $\cu(1)$, or the group
$Sp(1) \simeq SU(2)$ respectively. And $\pi : P \raw M$ will be the
corresponding $G$ principal (super)fibration, with $P$ the sphere $S^3$, the
supergroup manifold $UOSP(1,2)$, or the sphere $S^7$. The symbol $\IF$ will stand
for the vectors spaces underlying the field of complex numbers $\IC$, a complex
Grassmann algebra $C_L$ with $L$ generators or the field of quaternions $\IH$. 
We shall indicate with $\cb_{\IF}$ the algebra of $\IF$-valued smooth
functions on the total space $P$, while $\ca_{\IF}$ will be the algebra of
$\IF$-valued smooth functions on the base space $M$, thus
$\cb_{\IF}=:C^\infty(P,\IF)$ and $\ca_{\IF}=:C^\infty(M,\IF)$.

On $\IF$ there is a left action of the group $G$ and $C^\infty_{G}(P, \IF)$ will
denote the collection of corresponding equivariant maps:
\be
\varphi : P ~\raw~ \IF~, ~~\varphi(p\cdot w) = w^{-1}\cdot \varphi(p)~, 
\ee
with $\varphi \in C^\infty_{G}(P, \IF)$ and for any $p\in P$, $w\in G$; it
is a right module over $\ca_{\IF}$. It is well known (see for instance \cite{Tr})
that there is a module isomorphism between $C^\infty_{G}(P, \IF)$ and the right
$\ca_{\IF}$-module of sections $\Gamma^\infty(M, E)$ of the associated vector
bundle $E=P\times_G\IF$ over $M$. In the spirit of Serre-Swan's theorem \cite{Sw},
the module
$\Gamma^\infty(M, E)$ will be identified with the image in the trivial module
$(\ca_{\IF})^N$ of a projector $p\in \IM_N(\ca_{\IF})$, the latter being the
algebra of $N\times N$ matrices with entries in $\ca_{\IF}$, i.e. $\Gamma^\infty(M,
E) = p (\ca_{\IF})^N$. Since this projector is a rank $1$ (over $\IF$) it will be
written as 
\be\label{urpro}
p=\ket{\psi}\bra{\psi}~, 
\ee
with 
\be\label{urket}
\ket{\psi}=
\left(
\begin{array}{c}
\psi_1 \\ \vdots \\ \psi_N 
\end{array}
\right)~,
\ee
a specific vector-valued function on $P$, thus a specific element of
$(\cb_\IF)^N$, the components being functions $\psi_i \in \cb_{\IF}~, ~i =1, \dots,
N$. The vector-valued function will be normalized,
\be
\hs{\psi}{\psi}=1~,
\ee 
a fact implying that $p$ is a projector
\be
p^2 = \ket{\psi}\hs{\psi}{\psi}\bra{\psi} = p~, ~~~p^\dagger = p~,
\ee
with $^\dagger$ a suitable adjoint (see later). Furthermore, the normalization will
also imply that $p$ is of rank $1$ over $\IF$ because
\be\label{tragen}
tr_\IF ~p = \hs{\psi}{\psi} = 1~.
\ee
In fact, the right end side of (\ref{tragen}) is not the number $1$ but rather the
constant function $1$; then a normalized integration will yield the number $1$
as the value for the rank of the projector and of the associated vector bundle. The
transformation rule of the vector-valued function
$\ket{\psi}$ under the right action of an element $w \in G$ will be very simple,
being indeed just component-wise multiplication,
\be
\ket{\psi} \mapsto \ket{\psi^w} =: 
\left(
\begin{array}{c}
\psi_1 w \\ \vdots \\ \psi_N w
\end{array}
\right) =: \ket{\psi} w ~.
\ee
As a consequence, the projector $p$ will be invariant under the right action
of $G$,
\be
p \mapsto p^w = \ket{\psi^w}\bra{\psi^w} = \ket{\psi} w w^\dagger \bra{\psi} 
= p~,
\ee
being $w w^\dagger=1$. Thus, the entries of $p$ are functions on the base space
$M$, that is are elements of the algebra $\ca_\IF$ and $p\in \IM_N(\ca_{\IF})$, as
it should be.

To keep things distinct, we shall denote elements of $(\ca_{\IF})^N$ by the
symbol 
\be
\ket{\ket{f}}=
\left(
\begin{array}{c}
f_1 \\ \vdots \\ f_N 
\end{array}
\right)~,
\ee
with $f_1, \dots, f_N$, elements of $\ca_{\IF}$. The module isomorphism
between sections and equivariant maps will be explicitly  given by,
\bea\label{isoseem}
&& \Gamma^\infty(M,E) ~\lrw~ C^\infty_{G}(P, \IF)~, \nn \\
&& \sigma = p\ket{\ket{f}} ~\lrw~ \varphi_\sigma = 
\hs{\psi}{\ket{f}} =: \sum_{i=1}^N \bar{\psi}_i f_i~.
\eea
In fact, as we shall see, we shall first construct the equivariant maps, out
of them the projector and then the sections of the associated bundle.

Having the projector, we can define a canonical connection (also called the
Grassmann connection) on the module of sections by,
\bea\label{con}
&&\nabla =: p \circ d ~:~ \Gamma^\infty(M,E) ~\raw~ \Gamma^\infty(M,E)
\otimes_{\ca_{\IF}} \Omega^1(M, \IF)~, \nn \\
&&\nabla\sigma =: \nabla(p\ket{\ket{f}}) = p( \ket{\ket{df}} +
dp\ket{\ket{f}})~,
\eea
where we have used the explicit identification $\Gamma^\infty(M,
E) = p (\ca_{\IF})^N$. The corresponding
connection
$1$-form $A_\nabla \in End_{\cb_{\IF}}(C^\infty_{G}(P,\IC)) \otimes_{\cb_{\IF}}
\Omega^1(P, \IF)$ on the equivariant maps has a very simple expression in terms of 
the vector-valued function
$\ket{\psi}$. Indeed, for any section $\sigma \in \Gamma^\infty(M,E)$, by
using the isomorphism (\ref{isoseem}) and Leibniz rule we find,
\bea
\nabla \varphi_\sigma =: \varphi_{\nabla \sigma} 
&=& \hs{\psi}{(\ket{\ket{df}} + d(\ket{\psi}\bra{\psi})\ket{\ket{f}})} \nn \\
&=& \hs{\psi}{(\ket{\ket{df}} + \ket{\psi}\hs{d\psi}{\ket{f}}) +
\ket{d\psi}\hs{\psi}{\ket{f}}} \nn \\
&=& d \hs{\psi}{\ket{f}} + \hs{\psi}{d \psi} \hs{\psi}{\ket{f}} \nn \\
&=& (d + \hs{\psi}{d \psi}) \hs{\psi}{\ket{f}} \nn \\
&=& (d + \hs{\psi}{d \psi}) \varphi_\sigma~,
\eea
from which we get 
\be\label{confor}
A_\nabla = \hs{\psi}{d \psi}~.
\ee
This connection form is anti-hermitian, a consequence of the normalization
$\hs{\psi}{\psi}=1$:
\be
(A_\nabla)^\dagger =: \hs{d\psi}{\psi} = - \hs{\psi}{d \psi} = -A_\nabla~.
\ee

As for the curvature of the connection (\ref{con}), which is $\nabla^2$, it is
found to be,
\be\label{cur}
\nabla^2 = p (dp)^2~. 
\ee
We shall also need the operator $\nabla^2\circ\nabla^2$ which is readily found
to be
\be
\label{cursqu}
\nabla^2\circ\nabla^2 = p (dp)^4~.
\ee
By using a suitable trace, these two operators give the first and the second Chern
classes of the vector bundle (these are the only classes that we shall need) as
\cite{Co},
\bea\label{checla}
&&C_1(p) =: - {1 \over 2 \pi i} ~tr(\nabla^2) 
= - {1 \over 2 \pi i} ~tr (p (dp)^2)~, \nn \\ 
&&C_2(p) =: {1 \over 2}(-{1\over 2 \pi i})^2 tr ( \nabla^2\circ\nabla^2) 
= - {1\over 8 \pi^2 } ~tr (p (dp)^4)~.
\eea
When integrated over $M$, they will give the corresponding Chern numbers,
\be\label{chenum}
c_1(p) = \int_M C_1(p) ~, ~~~c_2(p) = \int_M C_2(p) ~.
\ee 

On the ket-valued function (\ref{urket}) there will also be a global {\it left}
action of a group of `unitaries' $SU = \{s ~|~ s s^\dagger = 1 \}$ (for the three
cases considered this group will be $SU(2)$, $UOSP(1,2)$ and $Sp(2)\simeq Spin(5)$
respectively) which preserves the normalization, 
\be
\ket{\psi} ~\mapsto~ \ket{\psi^s} = s \ket{\psi}~, 
~~~ \hs{\psi^s}{\psi^s} = 1~.
\ee
The corresponding transformed projector 
\be
p^s = \ket{\psi^s} \bra{\psi^s} = s \ket{\psi} \bra{\psi} s^\dagger = s p
s^\dagger~,
\ee
is equivalent to the starting one, the partial isometry being $v = s p$; indeed,
$vv^\dagger = p^s$ and $v^\dagger v = p$.  Furthermore, the connection $1$-form
is left invariant,
\be\label{invcon}
A_{\nabla^s} = \hs{\psi^s}{d \psi^s} = \bra{\psi}s^\dagger s\ket{d \psi} =
A_\nabla~.
\ee

To get new (in general gauge non-equivalent) connections one should act with
group elements which do not preserve the normalization. Thus let $g$ in some
group
$GL$ (for the three cases we shall study this group will be $GL(2; \IC)$,
$GL(1,2;C_L)$ and $GL(2;\IH)$ respectively) act on the ket-valued function
(\ref{urket}) by 
\be
\ket{\psi} ~\mapsto~ \ket{\psi^g} = 
{1 \over [\bra{\psi}g^\dagger g\ket{\psi}]^{{1 \over 2}}}
~g \ket{\psi} ~.
\ee
The corresponding transformed projector
\bea
p^g &=& \ket{\psi^g} \bra{\psi^g} = 
{1 \over \bra{\psi}g^\dagger g\ket{\psi}} ~g \ket{\psi} \bra{\psi} g^\dagger~,
\nn \\  
&=& 
{1 \over \bra{\psi}g^\dagger g\ket{\psi}} ~g p g^\dagger
\eea
is again equivalent to the starting one, the partial isometry being now,
\be
v = {1 \over [\bra{\psi}g^\dagger g\ket{\psi}]^{{1 \over 2}}} ~g p ~.
\ee
Indeed,
\bea
&& vv^\dagger = p^s~, \nn \\
&& v^\dagger v = {1 \over \bra{\psi}g^\dagger g\ket{\psi}} ~ p g^\dagger g p = 
{1 \over \bra{\psi}g^\dagger g\ket{\psi}} ~\ket{\psi}\bra{\psi} g^\dagger g
\ket{\psi}\bra{\psi} \nn \\
&& ~~~~~~ = \ket{\psi}\bra{\psi} = p~.
\eea
The associated connection $1$-form is readily found to be
\be\label{tracon}
A_{\nabla^g} =: \hs{\psi^g}{d \psi^g} = 
{1 \over 2 \bra{\psi}g^\dagger g\ket{\psi}} 
~[\bra{\psi}g^\dagger g\ket{d \psi} - \bra{d \psi}g^\dagger g\ket{\psi}]~.
\ee
Thus, if $g \in SU$, we get back the previous invariance of
connections (\ref{invcon}), while for 
$g \in GL$ modulo $SU$ we get new, gauge non-equivalent connections on the
$\IF$-line bundle over $M$ determined by the projector $p^g$, line bundle which is
(stable) isomorphic to the one determined by the projector $p$.
 
We end these general considerations with some remarks on the operation of
transposition of projectors. If we transpose the projector (\ref{urpro}) we still
get a projector,
\be
q =: p^{t} = \ket{\phi} \bra{\phi}~,
\ee
with the transposed ket-valued functions given by,
\be\label{bratra}
\ket{\phi}=: (\bra{\psi})^{t} = 
\left(
\begin{array}{c}
\bar{\psi}_1 \\ \vdots \\ \bar{\psi}_N 
\end{array}
\right)~. 
\ee
That $q$ is a projector ($q^2=q$) of rank $1$ (over $\IF$) are both consequences
of the normalization $ \hs{\phi}{\phi} = \hs{\psi}{\psi} = 1$. But it turns out
that the transposed projector is {\it not} equivalent to the starting one, the
corresponding topological charges differing in sign, a change in sign which comes
from the antisymmetry of the exterior product for forms. In the present paper this
will be shown only for the lowest values ($\pm 1$) of the charge while the general
case will be presented in \cite{La}. Thus, transposing of projectors yields an
isomorphism in $K$-theory which is not the identity map.

\sect{The Dirac Monopole}
\subsect{The $U(1)$ Bundle over $S^2$}

The $U(1)$ principal fibration $\pi : S^3 \raw S^2$ over the two
dimensional sphere is explicitly realized as follows.
The total space is 
\be
S^3 = \{(a,b) \in \IC^2 ~,~ |a|^2 + |b|^2 = 1\}~. 
\ee
with right $U(1)$-action 
\be\label{u1act}
S^3 \times U(1) ~\raw~ S^3~, ~~~(a,b) w = (aw, bw);
\ee
Clearly $|a w|^2 + |b w|^2 = |a|^2 + |b|^2 = 1$. 
The bundle projection $\pi : S^3 \raw S^2$ is just the Hopf projection and it is 
given by $\pi(a,b) =: (x_0, x_1, x_2)$, 
\bea\label{s2coord}
&& x_0 = |a|^2 - |b|^2 = -1 +2|a|^2 = 1 - 2|b|^2 ~, \nn \\ 
&& x_1 = a\bar{b} + b\bar{a} ~, \nn \\
&& x_2 = i(a\bar{b} - b\bar{a}) ~,
\eea
and one checks that $\sum_{\mu=0}^2 (x_\mu)^2 = (|a|^2 + |b|^2)^2 = 1$~. 
The inversion of (\ref{s2coord}) gives the basic ($\IC$-valued) invariant functions
on $S^3$,
\bea
&& |a|^2 = {1 \over 2} (1+x_0)~, \nn \\
&& |b|^2 = {1 \over 2} (1-x_0)~, \nn \\
&& a \bar{b} = {1 \over 2} (x_1 -i x_2)~,
\eea
a generic invariant (polynomial) function on $S^3$ being any function of the
previous variables. Later on we shall need the volume form of $S^2$ which turns out
to be
\be\label{vols2}
d (vol(S^2)) = x_0 dx_1 dx_2 + x_1 dx_2 dx_0 + x_2 dx_0 dx_1 = 2 i (da d\bar{a} +
db d\bar{b})~.
\ee

\subsect{The Bundle and the Projector for the Monopole}

We need the rank $1$ complex vector bundle associated with the defining left 
representation of $U(1)$ on $\IC$. The latter is just left complex
multiplication by $w \in U(1)$,
\be\label{lefu1}
U(1) \times \IC ~\raw~ \IC~, ~~~(w,c) \mapsto c' = w c ~.
\ee
The corresponding equivariant maps $\varphi : S^3 \raw \IC$ are of the form
\be\label{equmon}
\varphi(a,b) = \bar{a} f + \bar{b} g~,
\ee
with $f,g$ any two $\IC$-valued functions which are invariant under the
right action of $U(1)$ on $S^3$. Indeed,
\be\label{equmon1}
\varphi((a,b)w) = \bar{aw} f + \bar{bw} g = w^{-1} \varphi(a,b)~.
\ee
We shall think of $f,g$ as $\IC$-valued functions
on the base space $S^2$, namely elements of $\ca_{\IC} =: C^\infty(S^2, \IC)$. The
space $C^\infty_{U(1)}(S^3, \IC)$ of equivariant maps is a right module over the
(pull-back of) functions $\ca_{\IC}$.

Consider then, the vector-valued function,
\be\label{bramon}
\ket{\psi}=: 
\left(
\begin{array}{c}
a \\ b
\end{array}
\right)~,
\ee
which is normalized
\be\hs{\psi}{\psi}= \abs{a}^2 + \abs{b}^2 = 1~.
\ee
By using it we can construct a projector in $\IM_2(\ca_{\IC})$:
\be\label{promon}
p =: \ket{\psi} \bra{\psi} = 
\left(
\begin{array}{cc}
\abs{a}^2 & a\bar{b} \\ 
b\bar{a} & \abs{b}^2
\end{array}
\right) = 
{1 \over 2} \left(
\begin{array}{cc}
1 + x_0 & x_1 - i x_2 \\ 
x_1 + i x_2 & 1 - x_0
\end{array}
\right)~,
\ee
where we have used the definition (\ref{s2coord}) for the coordinates on $S^2$. It
is clear that $p$ is a projector,
\be
p^2 =: \ket{\psi} \hs{\psi}{\psi} \bra{\psi} 
= \ket{\psi} \bra{\psi} = p~, ~~~p^\dagger = p~.
\ee
Moreover, it is of rank $1$ over $\IC$ because its trace is the constant function
$1$,
\be
tr p = \hs{\psi}{\psi} = 1~.
\ee
The $U(1)$-action (\ref{u1act}) will transform the vector (\ref{bramon})
multiplicatively,
\be
\ket{\psi} ~\mapsto~ \ket{\psi^w} = 
\left(
\begin{array}{c}
a w \\ b w
\end{array}
\right) = \ket{\psi}w~, ~~~\forall ~w \in U(1)~.
\ee
As a consequence the projector $p$ is invariant, a fact which is also evident
from its explicit expression (\ref{promon}).

Thus, the right module of sections $\Gamma^\infty(S^2, E)$ of the associated
bundle is identified with the image of $p$ in $(\ca_{\IC})^2$ and the module
isomorphism between sections and equivariant maps is given by,
\bea
&& \Gamma^\infty(S^2,E) ~\lrw~ C^\infty_{U(1)}(S^3, \IC)~, \nn \\
&& \sigma = p 
\left(
\begin{array}{l}
f \\ g
\end{array}
\right) ~\lrw~ \varphi_\sigma = 
\bra{\psi} \left(
\begin{array}{l}
f \\ g
\end{array}
\right) = \bar{a} f + \bar{b} g ~, ~~~\forall ~f,g \in \ca_{\IC}~. 
\eea
It is obvious that this map is a module isomorphism.

The canonical connection associated with the projector,
\be
\nabla = p \circ d ~:~ \Gamma^\infty(S^2,E) ~\raw~ \Gamma^\infty(S^2,E)
\otimes_{\ca_{\IC}} \Omega^1(S^2, \IC),
\ee
has curvature given by
\be\label{moncur}
\nabla^2 = p (dp)^2 = \ket{\psi}\hs{d \psi}{d \psi} \bra{\psi}~. 
\ee
The associated Chern $2$-form is 
\be\label{moncf}
C_1(p) =: - {1 \over 2 \pi i} ~tr (p (dp)^2) = - {1 \over 2 \pi i} \hs{d \psi}{d
\psi} = {1 \over 2 \pi i} (da d\bar{a} +  db d\bar{b}) = -{1 \over 4 \pi} 
d(vol(S^2))~,
\ee
with corresponding first Chern number 
\be\label{moncn}
c_1(p) = \int_{S^2} C_1(p) = -{1 \over 4 \pi} \int_{S^2} d (vol(S^2)) = 
-{1 \over 4 \pi} 4 \pi = -1~.
\ee

\bigskip

By transposing the projector (\ref{promon}) we obtain an inequivalent projector,
\be
q =: p^{t} = \ket{\phi} \bra{\phi}~,
\ee
with the transposed ket vector given by,
\be\label{bramontra}
\ket{\phi}=: (\bra{\psi})^{t} = 
\left(
\begin{array}{c}
\bar{a} \\ \bar{b}
\end{array}
\right)~. 
\ee
We find that
\be\label{promontra}
q = 
\left(
\begin{array}{cc}
\abs{a}^2 & b\bar{a} \\ 
a\bar{b} & \abs{b}^2
\end{array}
\right) = 
{1 \over 2} \left(
\begin{array}{cc}
1 + x_0 & x_1 + i x_2 \\ 
x_1 - i x_2 & 1 - x_0
\end{array}
\right)~.
\ee
That $q$ is a projector ($q^2=q$), of rank $1$ ($tr q = 1$) are both consequences
of the normalization $ \hs{\phi}{\phi} = \abs{a}^2 + \abs{b}^2 = 1$.
The projector $q$ is obtained from $p$ by exchanging $a \raw \bar{a}$
and $b \raw \bar{b}$ which amounts to the exchange $x_2 \raw -x_2$~. It is then
clear that the corresponding Chern form and number are given by,
\bea
&& C_1(q) = - {1 \over 2 \pi i} (da d\bar{a} +  db d\bar{b}) 
= {1 \over 4 \pi} d (vol(S^2))~, \label{moncftra} \\
&& c_1(q) = \int_{S^2} C_1(q) = {1 \over 4 \pi} \int_{S^2} d (vol(S^2)) 
= 1~. \label{moncntra}
\eea
Having different topological charges the projectors $p$ and $q$ are clearly
inequivalent. This inequivalence is a manifestation of the fact that transposing
of projectors yields an isomorphism in the reduced group $\widetilde{K}(S^2)$,
which is not the identity map.

\bigskip

The connection $1$-form (\ref{confor}) associated with the projector $p$ is
given by  
\be\label{conformon}
A_\nabla = \hs{\psi}{d \psi} = \bar{a} da + \bar{b} db~.
\ee
This connection form is clearly anti-hermitian, so it is valued in $i \IR$, the
Lie algebra of $U(1)$. It coincides with the charge $-1$ monopole connection
form \cite{Ma,Tr}. Furthermore, the invariance (\ref{invcon}) states the
invariance of (\ref{conformon}) under left action of $SU(2)$. Gauge non-equivalent
connections are obtained by the formula (\ref{tracon}),
\be\label{traconmon}
A_{\nabla^g} =
{1 \over 2 \bra{\psi}g^\dagger g\ket{\psi}} 
~[\bra{\psi}g^\dagger g\ket{d \psi} - \bra{d \psi}g^\dagger g\ket{\psi}]~,
~~~
\ket{\psi}=: 
\left(
\begin{array}{c}
a \\ b
\end{array}
\right)~,
\ee
with $g \in GL(2;\IC)$ modulo $SU(2)$. Similar considerations hold for the
connection $1$-form associated with the monopole projector $q$.

\def\dia{^\diamond}
\def\edi{\eta\dia}
\sect{The Graded Monopole}
\subsect{The $\cu(1)$ Bundle over $S^{2,2}$}

The $\cu(1)$ principal fibration $\pi : UOSP(1,2) \raw S^{2,2}$ over the
$(2,2,)$-dimensional supersphere is explicitly realized as follows. 
The total space is the $(1,2,)$-dimensional supergroup $UOSP(1,2)$. Both
$UOSP(1,2)$ and $\cu(1)$ are described in more details in the Appendix. A
generic element
$s \in UOSP(1,2)$ can be parametrized as 
\be\label{supertotal}
s = \left(
\begin{array}{ccc}
1 + {1 \over 4}\eta \edi & - {1 \over 2}\eta & {1 \over 2}\edi \\
~&~&~\\
 -{1 \over 2}(a \edi - b\dia \eta) & ~a (1 - {1 \over 8}\eta \edi) ~
& ~-b\dia (1 - {1 \over 8}\eta \edi) ~ \\ 
 ~&~&~\\
~-{1 \over 2}(b \edi + a\dia \eta)~ & ~b (1 - {1 \over 8}\eta \edi) ~ 
& ~a\dia (1 - {1 \over 8}\eta \edi) ~
\end{array}
\right)~.
\ee
Here $a, b$ and $\eta$ are elements in a complexified Grassmann algebra
$C_L = B_L \otr \IC$ with $L$ generators ($B_L$ being obviously a real Grassmann
algebra) with the restriction $a, b \in (C_L)_0$ and $\eta \in (C_L)_1$.
Furthermore, elements of $UOSP(1,2)$ have superdeterminant equal to $1$ and this
give the condition
\be
a a\dia + b b\dia = Sdet(s) = 1~.
\ee
We recall that the integer $L$ is taken to be even and this assures \cite{RS} the
existence of an even graded involution 
\bea
\dia : C_L ~\raw~ C_L~, 
&& \abs{x\dia} = \abs{x}~, ~~~\forall ~x\in (C_L)_{\abs{x}}~, \nn \\ 
&& (c x)\dia = \bar{c} x\dia~, ~~~\forall ~c\in \IC~, ~x\in C_L ~,
\eea
which in addition verifies the properties 
\bea
&& (x y)\dia = x\dia ~y\dia~,  ~~~\forall ~x,y\in C_L~, \nn \\
&& x {\dia}{\dia} = (-1)^{\abs{x}}x~, ~~~\forall ~x\in (C_L)_{\abs{x}}~.
\eea

The structure supergroup of the fibration is $\cu(1)$, the Grassmann extension of
$U(1)$. It can be realized as follows
\be
\cu(1) = \{ w \in (C_L)_{0} ~|~ w w\dia = 1 \}.
\ee
Now, the Lie superalgebra of $UOSP(1,2)$ is generated by three even elements
${A_0, A_1, A_2}$ and two odd ones ${R_+, R_-}$ whose matrix
representation is given in (\ref{osp12}). Then, by embedding $\cu(1)$ in $UOSP(1,2)$
as 
\be
w ~\mapsto~ \left(
\begin{array}{ccc}
1 & 0 & 0 \\
0 & w & 0 \\ 
0 & 0 & w\dia~
\end{array}
\right)~,
\ee
we may think of $A_0$ as the generator of $\cu(1)$, i.e.
\be
\cu(1) \simeq \{ exp(\lambda A_0) ~|~ \lambda \in (C_L)_{0}~, ~\lambda\dia =
\lambda\}.
\ee
We let $\cu(1)$ act on the right on $UOSP(1,2)$. If we parametrize any 
$s \in UOSP(1,2)$ by $s = s(a,b,\eta)$, then this action can be represented as
follows,
\be\label{cu1act}
UOSP(1,2) \times \cu(1) ~\raw~ UOSP(1,2)~, ~~~(s, w) \mapsto 
s \cdot w = s(aw,bw,\eta w)~. 
\ee
This action leaves unchanged the superdeterminant
$Sdet(s\cdot w) = aw (aw)\dia + bw (bw)\dia = a a\dia + b b\dia =1$.

The bundle projection
\bea
&& \pi : UOSP(1,2) ~\raw~ S^{2,2} =: UOSP(1,2) / \cu(1)~, \nn \\
&& \pi(a,b, \eta) = (x_0, x_1, x_2, \xi_+, \xi_-)
\eea
can be given as the (co)-adjoint orbit through $A_0$. With $s^\dagger$
the adjoint of $s$ as given in (\ref{supadj}), one has that 
\be
\pi(s) =: s ({2 \over i} A_0) s^\dagger =: \sum_{k=0,1,2} x_k ({2 \over i} A_k) +
\sum_{\alpha=+,-} \xi_\alpha (2 R_\alpha)~.
\ee
Explicitly, 
\bea\label{ss2coord}
&& x_0 = (a a\dia - b b\dia)(1 - {1 \over 4}\eta \edi) = 
(-1 + 2a a\dia)(1 - {1 \over 4}\eta \edi) = 
(1 - 2b b\dia)(1 - {1 \over 4}\eta \edi)~, \nn \\  
&& x_1 = (a\bar{b} + b\bar{a})(1 - {1 \over 4}\eta \edi)~, \nn \\ 
&& x_2 = i(a\bar{b} - b\bar{a})(1 - {1 \over 4}\eta \edi)~, \nn \\
&& \xi_- = -{1 \over 2}(a \edi + \eta b\dia) ~, \nn \\
&& \xi_+ = {1 \over 2}(\eta a\dia - b \edi)~.
\eea
One sees directly that the $x_k$'s are even, $x_k \in (C_L)_{0}$, and `real', $x_k
\dia = x_k$, while the $\xi_\alpha$ are odd, $\xi_\alpha \in (C_L)_{1}$, and such
that $\xi_-{\dia} = \xi_+$ (and $\xi_+{\dia} = -\xi_-$). Furthermore,
\bea
\sum_{\mu=0}^2 (x_\mu)^2 + 2 \xi_- \xi_+ &=& 
(a a\dia + b b\dia)^2 (1 - {1 \over 2}\eta \edi) 
+ {1 \over 2}(a a\dia + b b\dia)\eta \edi \nn \\ 
&=& 1 ~.
\eea
Thus, the base space $S^{2,2}$ is a $(2,2)$-dimensional sphere in the superspace 
$B_L^{3,2}$. It turns out that $S^{2,2}$ is a De Witt supermanifold with {\it body}
the usual sphere $S^2$ in $\IR^3$ \cite{BBL}, a fact that we shall use later.
The inversion of (\ref{ss2coord}) gives the basic ($C_L$-valued) invariant
functions on $UOSP(1,2)$. Firstly, notice that 
\be
{1 \over 4}\eta \edi = \xi_-\xi_+~. 
\ee
Furthermore, 
\bea
&& a a\dia = {1 \over 2} [1 +  x_0 (1 + \xi_-\xi_+)] ~, \nn \\
&& b b\dia = {1 \over 2} [1 -  x_0 (1 + \xi_-\xi_+)] ~, \nn \\ 
&& a b\dia = {1 \over 2} (x_1 -i x_2)(1 + \xi_-\xi_+) ~, \nn \\
&& \eta a\dia = -(x_1 +i x_2)\xi_- + (1+x_0)\xi_+  ~, \nn \\
&& \eta b\dia = (x_1 -i x_2)\xi_+ - (1-x_0)\xi_-  ~,
\eea
a generic invariant (polynomial) function on $UOSP(1,2)$ being any function of the
previous variables. 

\subsect{The Bundle and the Projector for the Graded Monopole}

We need the rank $1$ (over $C_L$) vector bundle associated with the defining left 
representation of $\cu(1)$ on $C_L$. The latter is just left complex
multiplication by $w \in \cu(1)$,
\be\label{slefu1}
\cu(1) \times \IC ~\raw~ \IC~, ~~~(w,c) \mapsto c' = w c ~.
\ee
The corresponding equivariant maps $\varphi : UOSP(1,2) \raw C_L$ will be
written in the following form, 
\be\label{sequmon}
\varphi(\eta, a,b) = {1 \over 2}\edi h + a\dia (1 - {1 \over 8}\eta \edi) f +
b\dia (1 - {1 \over 8}\eta \edi) g~,
\ee
with $h,f,g$ any three $B_L$-valued functions which are invariant under the
right action of $\cu(1)$ on $UOSP(1,2)$ (the reason for the choice of the
additional invariant factor $(1 - {1 \over 8}\eta \edi)$ will be given later).
Indeed,
\bea\label{sequmon1}
\varphi((\eta, a,b)w) &=& {1 \over 2}(\eta w)\edi h 
+ (aw)\dia (1 - {1 \over 8}\eta \edi)f 
+ (bw)\dia (1 - {1 \over 8}\eta \edi) g \nn \\
&=& w^{-1}\varphi(\eta,a,b)~.
\eea
We shall think of $h,f,g$ as $C_L$-valued functions
on the base space $S^{2,2}$, namely elements of the algebra of
superfunctions $\ca_{C_L} =:C^\infty(S^{2,2},C_L)$ 
\footnote{We refer to \cite{BBH} for the definition of superfunctions where
they are denoted with $\cg$.}.  The space $C^\infty_{\cu(1)}(S^{2,2},C_L)$ of
equivariant maps is a right module over the (pull-back of) superfunctions
$\ca_{C_L}$.

Next, let us consider the ket-valued superfunction,
\be\label{sbramon}
\ket{\psi} =
\left(
\begin{array}{c}
-{1 \over 2}\eta \\
~\\ 
a (1 - {1 \over 8}\eta \edi) \\ 
~\\
b (1 - {1 \over 8}\eta \edi) 
\end{array}
\right)~,
\ee
whose associated bra is given by
\be
\bra{\psi}=: ({1 \over 2}\edi, ~a\dia (1 - {1 \over 8}\eta \edi), 
~b\dia (1 - {1 \over 8}\eta \edi))~.
\ee
They obey the relation
\bea
\hs{\psi}{\psi} &=& -{1 \over 4}\edi\eta + (a a\dia + b b\dia)
(1 - {1 \over 4}\eta \edi) \nn \\
&=& 1~.
\eea
As a consequence, we get a projector in $\IM_{2,1}(\ca_{C_L})$,
\bea\label{spromon}
p =: \ket{\psi} \bra{\psi} = 
\left(
\begin{array}{ccc}
- {1 \over 4}\eta \edi & - {1 \over 2}\eta a\dia & - {1 \over 2}\eta b\dia \\
~&~&\\
{1 \over 2} a \edi & a a\dia (1 - {1 \over 4}\eta \edi) & a b\dia (1 - {1
\over 4}\eta \edi) \\  
~&~& \\
{1 \over 2} b \edi & b a\dia (1 - {1 \over 4}\eta \edi) & b b\dia
(1 - {1 \over 4}\eta \edi)
\end{array}
\right) ~~~~~~~~~~ ~~~~~~~~~~~~~~~\nn \\ 
~ \nn \\
~ \nn \\
= {1 \over 2} \left(
\begin{array}{l}
2 \xi_+ \xi_- ~;~~~ (x_1 + i x_2)\xi_- - (1 + x_0)\xi_+ ~;~~~
 -(x_1 - i x_2)\xi_+ + (1 - x_0)\xi_- \\ ~\\
-(x_1 - i x_2)\xi_+ - (1 + x_0)\xi_- ~;~~~ 1 + x_0 + \xi_+ \xi_- ~;~~~
 x_1 - i x_2 \\ ~\\ 
-(x_1 + i x_2)\xi_- - (1 - x_0)\xi_+ ~;~~~~~~ x_1 + i x_2 ~;~~~~~ 
1 - x_0 + \xi_+ \xi_-
\end{array}
\right), 
\eea
where we have used the definition (\ref{ss2coord}) for the coordinates on
$S^{2,2}$. It is clear that $p$ is a projector,
\be
p^2 =: \ket{\psi} \hs{\psi}{\psi} \bra{\psi} 
= \ket{\psi} \bra{\psi} = p~, ~~~p^\dagger = p~.
\ee
And it is of (super-)rank $1$, because its supertrace is the constant function $1$,
\be
Str p =: (-1)(- {1 \over 4}\eta \edi) + (a a\dia + b b\dia) 
(1 - {1 \over 4}\eta \edi) = 1~,
\ee
with $Str$ denoting the supertrace of a supermatrix. The action (\ref{cu1act}) of
$\cu(1)$ will transform the vector (\ref{sbramon}) by 
\be
\ket{\psi} ~\mapsto~ \ket{\psi^w} = 
\left(
\begin{array}{c}
-{1 \over 2}\eta w\\
~\\ 
a w (1 - {1 \over 8}\eta \edi) \\ 
~\\
b w (1 - {1 \over 8}\eta \edi) 
\end{array}
\right)
= \ket{\psi}w~, ~~~\forall ~w \in \cu(1)~,
\ee
and as a consequence the projector $p$ is left invariant, a fact which is also
evident from its explicit form (\ref{spromon}). 

Thus, the right module of sections
$C^\infty(S^{2,2},E)$ of the associated bundle is identified with the image of $p$
in $(\ca_{C_L})^3$ and the module isomorphism between sections and equivariant
maps is given by,
\bea
C^\infty(S^{2,2},E) &\lrw& C^\infty_{\cu(1)}(S^{2,2},C_L)~, \nn \\
\sigma = p 
\left(
\begin{array}{l}
h \\ f \\ g
\end{array}
\right) &\lrw& \varphi_\sigma =
\bra{\psi} \left(
\begin{array}{l}
h \\ f \\ g
\end{array}
\right) \nn \\
&&~~~~ = {1 \over 2}\edi h + a\dia (1 - {1 \over 8}\eta \edi) f +
b\dia (1 - {1 \over 8}\eta \edi) g ~, 
\eea
for any $h, f,g \in \ca_{C_L}$.

The canonical connection associated with $p$,
\be
\nabla = p \circ d ~:~ C^\infty(S^{2,2},E) ~\raw~ C^\infty(S^{2,2},E)
\otimes_{\ca_{C_L}} \Omega^1(S^{2,2}, \IC),
\ee
has curvature given by
\be\label{smoncur}
\nabla^2 = p (dp)^2 = \ket{\psi}\hs{d \psi}{d \psi} \bra{\psi}~. 
\ee
The associated Chern $2$-form is found to be
\bea\label{smoncf}
C_1(p) &=:& - {1 \over 2 \pi i} ~Str (p (dp)^2) = - {1 \over 2 \pi i} \hs{d \psi}{d
\psi} \nn \\ &=& - {1 \over 2 \pi i} [-(da da\dia + db db\dia)(1 - {1 \over 4}\eta
\edi)
\nn \\
&& ~~~~~~~~~~~~~~~~~~~~-{1\over 4}  (a da\dia + b db\dia)(d\eta \edi + \eta d\edi)
 - {1\over 4} d\eta d\edi ]~, \nn \\
&=& - {1 \over 2 \pi i} [-(da da\dia + db db\dia)
%\nn \\ && ~~~~~~~~~~
-{1\over 4}  d(a\edi) d(\eta a\dia) - 
{1\over 4}  d(b\edi) d(\eta b\dia)]~.
\eea
By using the coordinates on $S^{2,2}$ the previous $2$-form results in
\bea
&& C_1(p) =: - {1 \over 4 \pi }(x_0 dx_1 dx_2 + x_1 dx_2 dx_0 + x_2 dx_0 dx_1)
(1 + 3 \xi_-\xi_+) \nn \\ 
&& ~~~~~ - {1 \over 4 \pi i}[(dx_1 -i dx_2)\xi_+ d\xi_+ - (dx_1 +i dx_2)\xi_-
d\xi_- +  dx_0(\xi_- d\xi_+  + \xi_+ d\xi_-) \nn \\ 
&& ~~~~~~~~~~~~~~~ +(x_1 -i x_2)d\xi_+ d\xi_+ - (x_1 +i x_2)d\xi_-d\xi_-
-2x_0d\xi_-d\xi_+ ]~.
\eea

Finally, to compute the corresponding first Chern number we need the Berezin
integral over the supermanifold $S^{2,2}$. This is a rather simple task due to the
fact that
$S^{2,2}$ is a De Witt supermanifold over the two-dimensional sphere $S^2$ in
$\IR^3$. Indeed, by using the natural morphism of forms ~$\widetilde{~} :
\Omega^2(S^{2,2}) \raw \Omega^2(S^2)$, the first Chern number yielded by
the superform $C_1(p)$ is computed by \cite{Br}
\be\label{smoncn}
c_1(p) =: Ber_{S^{2,2}} C_1(p) =: \int_{S^2} \widetilde{C_1(p)}~.
\ee
It is straightforward to find the projected form $\widetilde{C_1(p)}$. The bundle
projection $\Phi: S^{2,2} \raw S^2$ is explicitly realized in terms of the body
map,
\be
\Phi(x_0,x_1,x_2;\xi_-,\xi_+) ~\raw~ (\sigma(x_0),\sigma(x_1),\sigma(x_2))~.
\ee
Recall that fermionic variables do not have body. On the other side, by
denoting  $\sigma_i = \sigma(x_i), ~i=0,1,2$, the $\sigma_i$'s are cartesian
coordinates for the sphere $S^2$ in $\IR^3$ and obey the condition $(\sigma_0)^2 +
(\sigma_1)^2 + (\sigma_2)^2 =1$. The projected form $\widetilde{C_1(p)}$ is found
to be
\be
\widetilde{C_1(p)} = -{1 \over 4\pi}(\sigma_0 d\sigma_1 d\sigma_2 + \sigma_1
d\sigma_2 d\sigma_0 + \sigma_2 d\sigma_0 d\sigma_1) = -{1 \over 4\pi} vol(S^2)~.
\ee
As a consequence
\be
c_1(p) = Ber_{S^{2,2}} C_1(p) = -{1 \over 4 \pi} \int_{S^2} d (vol(S^2)) = 
-{1 \over 4 \pi} 4 \pi = -1~.
\ee

\bigskip

By (super)transposing the projector (\ref{spromon}) we obtain an inequivalent
projector,
\be\label{spromontra}
q =: p^{st} = \ket{\phi} \bra{\phi}~,
\ee
with the (super)transposed ket and bra vectors given by,
\bea\label{sbramontra}
&& \ket{\phi} =: (\bra{\psi})^{st} = 
\left(
\begin{array}{c}
{1 \over 2}\edi \\
~\\ 
a\dia (1 - {1 \over 8}\eta \edi) \\ 
~\\ 
b\dia (1 - {1 \over 8}\eta \edi) 
\end{array}
\right)~, \nn \\
&&~ \nn \\
&& \bra{\phi}=: (\ket{\psi})^{st} = ({1 \over 2}\eta, ~a(1 - {1 \over 8}\eta
\edi), ~b (1 - {1 \over 8}\eta \edi))~.
\eea
Explicitly, we find that
\bea
q = \left(
\begin{array}{ccc}
- {1 \over 4}\eta \edi & {1 \over 2}a \edi & {1 \over 2}b \edi \\
~&~&\\
{1 \over 2} a\dia \eta & a a\dia (1 - {1 \over 4}\eta \edi) & b a\dia (1 - {1
\over 4}\eta \edi) \\  
~&~&\\
{1 \over 2} b\dia \eta & a b\dia (1 - {1 \over 4}\eta \edi) & b b\dia
(1 - {1 \over 4}\eta \edi)
\end{array}
\right) ~~~~~~~~~~ ~~~~~~~~~~~~~~~ ~~~~~~~~~~~~~~~ ~~~ \nn \\ 
~ \nn \\
~ \nn \\
= {1 \over 2} \left(
\begin{array}{l}
2 \xi_+ \xi_- ~;~~~ -(x_1 - i x_2)\xi_+ - (1 + x_0)\xi_- ~;~~~
 -(x_1 + i x_2)\xi_- - (1 - x_0)\xi_+ \\ ~\\
-(x_1 + i x_2)\xi_- + (1 + x_0)\xi_+ ~;~~~ 1 + x_0 + \xi_+ \xi_- ~;~~~
 x_1 + i x_2 \\ ~\\ 
(x_1 - i x_2)\xi_+ - (1 - x_0)\xi_- ~;~~~~~~ x_1 - i x_2 ~;~~~~~ 
1 - x_0 + \xi_+ \xi_-
\end{array}
\right).
\eea
That $q$ is a projector ($q^2=q$), of rank $1$ ($Str q = 1$) are both consequences
of the normalization
\bea
\hs{\phi}{\phi} &=& {1 \over 4}\eta\edi + (a a\dia + b b\dia)
(1 - {1 \over 4}\eta \edi) \nn \\
&=& 1~.
\eea
We see that the transposed projector $q$ is obtained from $p$ by exchanging 
$a \lrw a\dia, b \lrw b\dia$ and $\eta \raw -\edi, ~\edi \raw \eta$. This 
amounts to the exchange of coordinates $x_2 \raw -x_2$ and $\xi_- \raw -\xi_+,
~\xi_+ \raw \xi_-$~. It is than clear that the corresponding Chern form and number
are given by,
\bea
&& C_1(q) = - C_1(p)~, \label{smoncftra} \\
&& c_1(q) = -c_1(p) = 1~. \label{smoncntra}
\eea
Having different topological charges the projectors $p$ and $q$ are clearly
inequivalent. Again, this inequivalence is a manifestation of the fact that
super-transposing of projectors yields an isomorphism in the reduced group
$\widetilde{K}(S^{2,2})$, which is not the identity map.

The connection $1$-form (\ref{confor}) associated with the projector $p$ is given
by  
\be\label{conforsupmon}
A_\nabla = \hs{\psi}{d \psi} = (a\dia da  + b\dia db)
(1 - {1 \over 4}\eta\edi) - {1 \over 8}(\edi d\eta  + \eta d\edi)~.
\ee
This connection form is clearly anti-hermitian, so it is valued in the Lie algebra
of $\cu(1)$. It coincides with the charge $-1$ graded monopole connection form
found in \cite{LM}. 
Furthermore, the invariance (\ref{invcon}) states the
invariance of (\ref{conforsupmon}) under left action of $UOSP(1,2)$. Gauge
non-equivalent connections are obtained by the formula (\ref{tracon})
\be\label{traconsmon}
A_{\nabla^g} =
{1 \over 2 \bra{\psi}g^\dagger g\ket{\psi}} 
~[\bra{\psi}g^\dagger g\ket{d \psi} - \bra{d \psi}g^\dagger g\ket{\psi}]~,
~~~
\ket{\psi} =
\left(
\begin{array}{c}
-{1 \over 2}\eta \\
~\\ 
a (1 - {1 \over 8}\eta \edi) \\ 
~\\
b (1 - {1 \over 8}\eta \edi) 
\end{array}
\right)~,
\ee
with $g \in GL(1,2;C_L)$ modulo $UOSP(1,2)$. Similar considerations hold for the
connection $1$-form associated with the monopole projector $q$.

\sect{The BPST Instanton}
\subsect{The $SU(2)$ Bundle over $S^4$}

The instanton could as well be called the quaternionic monopole due to the
strict similarity obtained when one trades complex numbers with quaternions. So we
shall start by giving a few fundamentals on quaternions $\IH$. For their basis we
shall use the symbols
$(1, i, j, k)$ with relations $i^2=-1, ij =k$, etc., so that any quaternion
$q\in\IH$ is written as
\be
q = r_0 + r_1 i + r_2 j + r_3 k~,
\ee
with $r_0, r_1, r_2, r_3$ real coefficients. 
The isomorphism $\IH \simeq \IC^2$ is realized as,
\bea\label{isonc2}
q &=& v_1 + v_2 j~, ~~~ v_1 = r_0 + r_1 i~, ~v_2 = r_2 + r_3 i \nn \\
\bar{q} &=& \bar{v}_1 -j \bar{v}_2 ~, ~~~ v j = j \bar{v}~, ~~~~~\forall ~v\in\IC
\eea
The quaternionic multiplication of $q = v_1 + v_2 j$ on the right by 
$w = w_1 + w_2 j$ is 
\be
q' =: qw = (v_1 w_1 - v_2 \bar{w}_2) + (v_1 w_2 + v_2 \bar{w}_1) j~,
\ee
giving a right matrix multiplication on $\IC^2$ via the isomorphism (\ref{isonc2})
\be\label{spsu}
\begin{array}{l}
(v'_1, v'_2) = (v_1, v_2) \\
~
\end{array}
\left(
\begin{array}{cc}
w_1 & w_2 \\ -\bar{w}_2 & \bar{w}_1  
\end{array}
\right)~.
\ee
In particular, if $w \in Sp(1)$, namely $w \bar{w} = 1 = |w_1|^2 + |w_2|^2$, 
the action (\ref{spsu}), provides the group isomorphism $Sp(1) \simeq SU(2)$,
\be
w = w_1 + w_2 j \mapsto 
\left(
\begin{array}{cc}
w_1 & w_2 \\ -\bar{w}_2 & \bar{w}_1  
\end{array}
\right)~.
\ee

The $SU(2) \simeq Sp(1)$ principal fibration $\pi : S^7 \raw S^4$ over the four
dimensional sphere is explicitly realized as follows.
The total space is 
\be
S^7 = \{(a,b) \in \IH^2 ~,~ |a|^2 + |b|^2 = 1\}~. 
\ee
with right action 
\be\label{sp1act}
S^7 \times Sp(1) ~\raw~ S^7~, ~~~(a,b) w = (aw, bw);
\ee
Clearly $|a w|^2 + |b w|^2 = |a|^2 + |b|^2 = 1$. 
The bundle projection $\pi : S^7 \raw S^4$ is just the Hopf projection and it
can be explicitly given as $\pi(a,b) = (x_0, x_1, x_2, x_3, x_4)$,
\bea\label{s4coord}
&& x_0 = |a|^2 - |b|^2 = -1 +2|a|^2 = 1 - 2|b|^2 ~, \nn \\ 
&& \xi = a\bar{b} - b\bar{a} =: x_1 i + x_2 j + x_3 k = -\bar{\xi} ~, \nn \\
&& x_4 = a\bar{b} + b\bar{a} ~.
\eea
One checks that $\sum_{\mu=0}^4 (x_\mu)^2 = (|a|^2 + |b|^2)^2 = 1$~.  
The inversion of (\ref{s4coord}) gives the basic ($\IH$-valued) invariant functions
on $S^7$,
\bea
&& |a|^2 = {1 \over 2} (1+x_0)~, \nn \\
&& |b|^2 = {1 \over 2} (1-x_0)~, \nn \\
&&  a \bar{b} = {1 \over 2} (x_4 + \xi)~, 
\eea
a generic invariant (polynomial) function on $S^7$ being any function of the
previous variables.

\subsect{The Bundle and the Projector for the Instanton}

We need the rank $2$ complex vector bundle associated with the defining left 
representation of $SU(2)$ on $\IC^2$. We shall realize it as the rank $1$
quaternionic vector bundle associated with the defining left representation of
$Sp(1)$ on $\IH$. For this we need a different identification $\IH \simeq
\IC^2$ from the one in (\ref{isonc2}), (a left identification)
\be\label{isonc3}
\l = \l_1 - j \l_2~, ~~~ \l_1 = r_0 + r_1 i~, ~\l_2 = r_2 + r_3 i~.
\ee
The quaternionic multiplication of $\l = \l_1 - j \l_2$
on the left by $w = w_1 + w_2 j \in Sp(1)$ is
\be
\l' =: w \l = (w_1 \l_1 + w_2 \l_2) -j (-\bar{w}_2 \l_1 + \bar{w}_1 \l_2) ~,
\ee
giving the left matrix multiplication of $SU(2)$ on $\IC^2$
\be\label{leftsu}
\left( 
\begin{array}{l}
\l'_1 \\ 
\l'_2 
\end{array}
\right) = 
\left(
\begin{array}{cc}
w_1 & w_2 \\ -\bar{w}_2 & \bar{w}_1  
\end{array}
\right) ~
\left( 
\begin{array}{l}
\l_1 \\ 
\l_2 
\end{array}
\right) ~.
\ee

The corresponding equivariant maps $\varphi : S^7 \raw \IH$ are of the form
\be\label{equins}
\varphi(a,b) = \bar{a} f + \bar{b} g~,
\ee
with $f,g$ any two $\IH$-valued functions which are invariant under the
right action of $Sp(1)$ on $S^7$. Indeed,
\be\label{equins1}
\varphi((a,b)w) = \bar{aw} f + \bar{bw} g = w^{-1} \varphi(a,b)~.
\ee
We shall think of $f,g$ as $\IH$-valued functions
on the base space $S^4$, namely elements of $\ca_\IH =: C^\infty(S^4, \IH)$. The
space $C^\infty_{Sp(1)}(S^7, \IH)$ of equivariant maps is a right module over the
(pull-back of) functions $\ca_\IH$.

Next, let us consider the ket-valued function,
\be\label{brains}
\ket{\psi}=: 
\left(
\begin{array}{c}
a \\ b
\end{array}
\right)~,
\ee
which satisfies
\be
\hs{\psi}{\psi}= \abs{a}^2 + \abs{b}^2 = 1~.
\ee
As before, we get a projector in $\IM_2(\ca_{\IH})$ by,
\be\label{proins}
p =: \ket{\psi} \bra{\psi} = 
\left(
\begin{array}{cc}
\abs{a}^2 & a\bar{b} \\ 
b\bar{a} & \abs{b}^2
\end{array}
\right) = 
{1 \over 2} \left(
\begin{array}{cc}
1 + x_0 & x_4 + \xi \\ 
x_4 - \xi & 1 - x_0
\end{array}
\right)~,
\ee
where we have used the definition (\ref{s4coord}) for the coordinates on $S^4$. It
is clear that $p$ is a projector,
\be
p^2 =: \ket{\psi} \hs{\psi}{\psi} \bra{\psi} 
= \ket{\psi} \bra{\psi} = p~, ~~~p^\dagger = p~.
\ee
Moreover, it is of rank $1$ over $\IH$ because its trace is the constant function
$1$,
\be
tr p = \hs{\psi}{\psi} = 1~.
\ee
The $Sp(1)$-action (\ref{sp1act}) will transform the vector (\ref{brains})
multiplicatively,
\be
\ket{\psi} ~\mapsto~ \ket{\psi^w} = 
\left(
\begin{array}{c}
a w \\ b w
\end{array}
\right) = \ket{\psi}w~, ~~~\forall ~w \in Sp(1)~,
\ee
while the projector $p$ remains unchanged, a fact which is also obvious from the
explicitly expression in (\ref{proins}).

Thus, the right module of sections $\Gamma^\infty(S^4,E)$ of the associated
bundle is identified with the image of $p$ in $(\ca_\IH)^2$ and the module
isomorphism between sections and equivariant maps is given by,
\bea
&& \Gamma^\infty(S^4,E) ~\lrw~ C^\infty_{Sp(1)}(S^7, \IH)~, \nn \\
&& \sigma = p 
\left(
\begin{array}{l}
f \\ g
\end{array}
\right) ~\lrw~ \varphi_\sigma =
\bra{\psi} \left(
\begin{array}{l}
f \\ g
\end{array}
\right) = \bar{a} f + \bar{b} g ~~~\forall ~f,g \in \ca_{\IH}~.
\eea

The canonical connection associated with the projector,
\be
\nabla = p \circ d ~:~ \Gamma^\infty(S^4,E) ~\raw~ \Gamma^\infty(S^4,E)
\otimes_{\ca_{\IH}} \Omega^1(S^4, \IH),
\ee
has curvature given by
\be\label{inscur}
\nabla^2 = p (dp)^2 = \ket{\psi}\hs{\psi}{d \psi}\hs{\psi}{d \psi} \bra{\psi} +
\ket{\psi}\hs{d \psi}{d \psi} \bra{\psi}~. 
\ee
Notice that, contrary to what happens for the monopole, the first term in
(\ref{inscur}) does not vanish since $\hs{\psi}{d \psi}$ is a quaternion-valued
$1$-form 

The associated Chern $2$-form and $4$-form are given respectively by 
\bea\label{inscf}
&&C_1(p) =: - {1 \over 2 \pi i} ~tr (p (dp)^2)~, \nn \\
&&C_2(p) =: - {1 \over 8 \pi^2 } ~[ tr (p (dp)^4) - C_1(p)C_1(p) ]~.
\eea
Now, in (\ref{inscf}), the trace $tr$ is really the tensor product of an ordinary
matrix trace with a trace $tr_{\IH}$ on $\IH$. By ciclicity $tr_{\IH}$ must
vanish on the imaginary quaternions. Indeed, for instance, $tr_{\IH}(i) =
tr_{\IH}(jk) = tr_{\IH}(jk) = - tr_{\IH}(i)$. Furthermore, we normalize it so that
\be\label{trah}
tr_{\IH}(1_{\IH}) = 2~;
\ee
this is motivated by the fact that a quaternion is a $2 \times 2$ matrix with
complex entries.

It turns out that the $2$-form $p (dp)^2$ is valued in the pure imaginary
quaternions. As a consequence its trace vanishes so we may conclude that
\be
C_1(p) = 0~.
\ee
As for the second Chern class, a straightforward calculation shows that,
\bea\label{inscf2}
C_2(p) &=& -{1 \over 32\pi^2} [(x_0 dx_4 - x_4 dx_0) (d\xi)^3 +
3 dx_0 dx_4 ~\xi ~(d\xi)^2] \nn \\
&=& -{3 \over 8\pi^2} [x_0 dx_1 dx_2 dx_3 dx_4 + x_0 dx_1 dx_2 dx_3 dx_4
\nn \\
& & ~~~~~~~~~~ + x_0 dx_1 dx_2 dx_3 dx_4 + x_0 dx_1 dx_2 dx_3 dx_4 
+ x_0 dx_1 dx_2 dx_3 dx_4 ] \nn \\
&=& -{3 \over 8\pi^2} ~d (vol(S^4))~.
\eea
The second Chern number is then given by
\be\label{inscn2}
c_2(p) = \int_{S^4} C_2(p) = -{3 \over 8\pi^2} \int_{S^4} d (vol(S^4)) = 
-{3 \over 8\pi^2} {8 \over 3} \pi^2 = -1~.
\ee

\bigskip

By transposing the  projector (\ref{proins}) we obtain an inequivalent projector, 
\be\label{proinstra}
q =: p^{t} = 
\left(
\begin{array}{cc}
\abs{a}^2 & b\bar{a} \\ 
a\bar{b} & \abs{b}^2
\end{array}
\right) = 
{1 \over 2} \left(
\begin{array}{cc}
1 + x_0 & x_4 - \xi \\ 
x_4 + \xi & 1 - x_0
\end{array}
\right)~.
\ee
In order to express this projector as a ket-bra in the way used so far, we need
to pay extra care to the noncommutativity of quaternions. Therefore, we introduce
the {\it right} ket vector $\ket{\phi}_R$ to be the {\it row} defined by
\be\label{brainstra}
\ket{\phi}_R=: (\ket{\psi})^{t} = (a, b)_R~. 
\ee
Then, contrary to what we have been doing so far, we multiply from {\it right} to
{\it left} in the expression,
\be
\begin{array}{l}\ket{\phi}_{R~R}\!\bra{\phi} = (a, b)_R \\
~
\end{array}
\!\!\left(
\begin{array}{l}
\bar{a} \\ \bar{b}
\end{array}
\right)_R ~.
\ee
With this convention we get the expression (\ref{proinstra}). Thus, we can write
\be
q = \ket{\phi}_{R~R}\!\bra{\phi}~,
\ee
and that $q$ is a projector ($q^2=q$), of rank $1$ ($tr q = 1$) are both
consequences of the normalization $ {_R}\!\hs{\phi}{\phi}_R = 
\abs{a}^2 + \abs{b}^2 = 1$. It is worth noticing that the corresponding bundle
of sections is realized as the {\it left} $(\ca_\IH)$-module $(\ca_\IH)^2 q$. The
transposed projector $q$ is obtained from $p$ by exchanging $\xi \raw -\xi$~. It is
than clear that the first Chern form still vanish, while, from the first equality
in (\ref{inscf2}), the second Chern form and the corresponding number are given by,
\bea
C_2(q) &=& {1 \over 32\pi^2} [(x_0 dx_4 - x_4 dx_0) (d\xi)^3 +
3 dx_0 dx_4 ~\xi ~(d\xi)^2] \nn \\
&=& {3 \over 8\pi^2} [x_0 dx_1 dx_2 dx_3 dx_4 + x_0 dx_1 dx_2 dx_3 dx_4
\nn \\
& & ~~~~~~~~~~ + x_0 dx_1 dx_2 dx_3 dx_4 + x_0 dx_1 dx_2 dx_3 dx_4 
+ x_0 dx_1 dx_2 dx_3 dx_4 ] \nn \\
&=& {3 \over 8\pi^2} ~d (vol(S^4))~. \label{instracf2} \\
&~& \nn \\
c_2(q) &=& \int_{S^4} C_2(q) = {3 \over 8\pi^2} \int_{S^4} d (vol(S^4)) = 
{3 \over 8\pi^2} {8 \over 3} \pi^2 = 1~. \label{instracn2}
\eea
Having different topological charges the projectors $p$ and $q$ are clearly
inequivalent. As mentioned already, this inequivalence is a manifestation of the
fact that transposing  projectors yields an isomorphism in the reduced group
$\widetilde{K}(S^4)$, which is not the identity map.

\bigskip

As for the connection $1$-form  (\ref{confor}) associated with the projector $p$,
it is given by  
\be\label{conforins}
A_\nabla = \hs{\psi}{d \psi} = \bar{a} da + \bar{b} db~.
\ee
This connection form is clearly anti-hermitian, so it is valued in the `purely
imaginary' quaternions which can be identified with the Lie algebra $sp(1)
\simeq su(2)$. It coincides with the charge $-1$ instanton
connection form \cite{At,Tr}. 
Furthermore, the invariance (\ref{invcon}) states the
invariance of (\ref{conforins}) under left action of $Sp(2) \simeq Spin(5)$. Gauge
non-equivalent connections are obtained by the formula (\ref{tracon}),
\be\label{traconins}
A_{\nabla^g} =
{1 \over 2 \bra{\psi}g^\dagger g\ket{\psi}} 
~[\bra{\psi}g^\dagger g\ket{d \psi} - \bra{d \psi}g^\dagger g\ket{\psi}]~,
~~~
\ket{\psi}=: 
\left(
\begin{array}{c}
a \\ b
\end{array}
\right)~,
\ee
with $g \in GL(2;\IH)$ modulo $Sp(2)$. However, not any transformed connection
deserves to be called an instanton since, in general, it needs not be (anti)-self
dual. Since the duality equations on $S^4$ are conformally invariant, it follows
that only conformal transformations will convert the  (anti)-instanton
(\ref{conforins}) in some other instanton. Now, it turns out that the proper (i.e.
orientation preserving) conformal group of $S^4$ is $SL(2;\IH)$ whose action, when
projected on $S^4$ reduces to  fractional linear transformations of a homogeneous
quaternionic variable (we recall that $S^4$ is naturally identifiable with the
quaternionic projective space
$\IH P^1$) \cite{At}. This action is the quaternionic analogue of the fact that the
the proper conformal group of $S^2$ is $SL(2;\IC)$ whose action, when projected on
$S^2$ reduces to fractional linear transformations of a homogeneous complex
variable, the sphere $S^2$ being naturally identifiable with the complex projective
space $\IC P^1$. Thus, with $g\in SL(2;\IH)$ modulo $Sp(2)$, the connection $1$-form
in (\ref{traconins}) represents an anti-instanton which is not gauge equivalent to
the starting anti-instanton (\ref{conforins}). Since $dim_{\IR}(SL(2;\IH)) -
dim_{\IR}(Sp(2)) = 15 - 10 = 5$, we get a $5$-parameter family of anti-instantons. 
Of course, the described procedure is nothing but the ADHM construction of
(anti)-instantons \cite{At}.
Similar considerations hold for the connection $1$-form associated with the
instanton projector $q$.

\bigskip\bigskip

\noindent
{\bf Acknowledgments}. This work was motivated by conversations with P. Hajac.
And I am grateful to him as well as to S. Winitzki for several useful
discussions and suggestions.

\bigskip\bigskip\bigskip

\noindent
{\em Note added}. After this paper appeared on the math-ph archive, K. Fujii made
me aware of his work \cite{Fu} where, by using Clifford algebras and stereographic
projections, he constructs (but does not deconstruct!) the charge $+1$
projectors on any even sphere $S^{2m}$.

\bigskip\bigskip\bigskip

%\vfill\eject
\def\dia{^\diamond}
\def\edi{\eta\dia}
\appendix
\sect{The Supergroup $UOSP(1,2)$}
We shall describe the basic facts about the supergroup  $UOSP(1,2)$ that we need
in this paper while referring to \cite{BT} for additional details.
Some of our notation differ from the one used in \cite{BT}.

With $B_L = (B_L)_0 + (B_L)_1$ we shall indicate a real Grassmann algebra with $L$
generators. Let $osp(1,2)$ be the Lie $B_L$ superalgebra of dimension $(3,2)$ with
even generators
${A_0, A_1, A_2}$ and odd generator ${R_+, R_-}$, explicitly given in matrix
representation by
\bea\label{osp12}
&&A_0 = {i \over 2}
\left(
\begin{array}{ccc}
0 & 0 & 0 \\
0 & 1 & 0 \\ 
0 & 0 & -1 
\end{array}
\right)~, ~~~
A_1 = {i \over 2}
\left(
\begin{array}{ccc}
0 & 0 & 0 \\
0 & 0 & 1 \\ 
0 & 1 & 0 
\end{array}
\right)~,
~~~
A_1 = {i \over 2}
\left(
\begin{array}{ccc}
0 & 0 & 0 \\
0 & 0 & -i \\ 
0 & i & 0 
\end{array}
\right)~, \nn \\
&&R_+ = {1 \over 2}
\left(
\begin{array}{ccc}
0 & -1 & 0 \\
0 & 0 & 0 \\ 
-1 & 0 & 0 
\end{array}
\right)~, ~~~
R_- = {1 \over 2}
\left(
\begin{array}{ccc}
0 & 0 & 1 \\
-1 & 0 & 0 \\ 
0 & 0 & 0 
\end{array}
\right)~.
\eea
Thus, a generic element $X\in osp(1,2)$ is written as $X = \sum_{k=0,1,2}
a_k A_k + \sum_{\alpha=+,-} \eta_\alpha R_\alpha$ with $a_k \in (B_L)_0,
~\eta_\alpha \in (B_L)_1$.

If the integer $L$ is taken to be even, on the complexification $C_L = B_L \otr
\IC$ there exists \cite{RS} an even graded involution $\dia : C_L \raw C_L$ which
satisfies the following properties,
\be
(x y)\dia = x\dia ~y\dia~,  ~~~\forall ~x,y\in C_L~, ~~~
x {\dia}{\dia} = (-1)^{\abs{x}}x~, ~~~\forall ~x\in (C_L)_{\abs{x}}~.
\ee
Next one introduces the Lie $C_L$ superalgebra $C_L \otr osp(1,2)$ and defines
the superalgebra $uosp(1,2)$ to be the `real' subalgebra made of elements of the
form
\be
X = \sum_{k=0,1,2} a_k A_k + \eta R_+ +\edi R_-~, ~~~a_k \in (C_L)_0~, ~a_k\dia
= a_k~, ~~\eta\in (C_L)_1~.
\ee
Indeed, one introduces an adjoint operation $^\dagger$ which is defined on the bases
(\ref{osp12}) as
\be
A_i^\dagger = - A_i~, ~~i = 0,1,2~; ~~R_+^\dagger = - R_-~, 
~R_-^\dagger = R_+~,
\ee
and is extended to the whole of $C_L \otr osp(1,2)$ by using the involution $\dia$.
Then, the superalgebra 
$uosp(1,2)$ is identified  as the collection of `anti-hermitian' elements
\be
uosp(1,2) = \{X \in C_L \otr osp(1,2) ~|~ X^\dagger = - X\}~.
\ee
The superalgebra $uosp(1,2)$ is the analogue of the compact real form of $C_L \otr
osp(1,2)$. 

Finally, the supergroup $UOSP(1,2)$ is defined to be the exponential map of
$uosp(1,2)$,
\be
UOSP(1,2) =: \{exp(X) ~|~ X \in uosp(1,2)\}~.
\ee
A generic element $s \in UOSP(1,2)$ can be presented as the product of
one-parameter subgroups,
\bea\label{s1ps}
&& s = u \xi ~, \nn \\
&& u = exp(a_0 A_0)exp(a_1 A_1)exp(a_1 A_1)~, ~~~a_k\dia = a_k \in (C_L)_0~, 
\nn \\
&& \xi = exp(\eta R_+ +\edi R_-)~, ~~~\eta \in (C_L)_0~.
\eea
Explicitly,
\be\label{sgroup}
s = \left(
\begin{array}{ccc}
1 + {1 \over 4}\eta \edi & - {1 \over 2}\eta & {1 \over 2}\edi \\
~&~&~\\
 -{1 \over 2}(a \edi - b\dia \eta) & ~a (1 - {1 \over 8}\eta \edi) ~
& ~-b\dia (1 - {1 \over 8}\eta \edi) ~ \\ 
 ~&~&~\\
~-{1 \over 2}(b \edi + a\dia \eta)~ & ~b (1 - {1 \over 8}\eta \edi) ~ 
& ~a\dia (1 - {1 \over 8}\eta \edi) ~
\end{array}
\right)~.
\ee
By using (\ref{s1ps}) one also finds the adjoint of any element to be
\bea\label{supadj}
&& s^\dagger =: \xi^\dagger u^\dagger \nn \\
&& ~ \nn \\
&& ~~~ 
= \left(
\begin{array}{ccc}
1 + {1 \over 4}\eta \edi & ~{1 \over 2}(a\dia \eta + b \edi)~  
& ~{1 \over 2}(b\dia \eta - a \edi)~ \\ 
~&~&~\\
{1 \over 2}\edi & ~a\dia (1 - {1 \over 8}\eta \edi)~
& ~b\dia (1 - {1 \over 8}\eta \edi)~ \\ 
~&~&~\\
 {1
\over 2}\eta & ~-b (1 - {1 \over 8}\eta \edi)~  & ~a(1 - {1 \over 8}\eta \edi)~
\end{array}
\right)~.
\eea

We have also used the one-parameter subgroup of $UOSP(1,2)$ generated by $A_0$, 
\be
\cu(1) \simeq \{ exp(\lambda A_0) ~|~ \lambda \in (C_L)_{0}~, ~\lambda\dia =
\lambda\}.
\ee
A generic element $w \in \cu(1)$ is written as
\be
w = \left( \begin{array}{ccc}
1 & 0 & 0 \\
0 & e^{{i \over 2} \lambda} & 0 \\ 
0 & 0 & e^{-{i \over 2} \lambda}
\end{array}
\right)~.
\ee

\bigskip\bigskip\bigskip

%\vfill\eject
\bibliographystyle{unsrt}

\end{document}